# Two-dimensional spatial coherence of excitons in semicrystalline polymeric semiconductors: The effect of molecular weight

Francis Paquin,[1] Hajime Yamagata,[2] Nicholas J. Hestand[2], Maciej Sakowicz,[1] Nicolas Bérubé,[1] Michel Côté,[1] Luke X. Reynolds,[3] Saif A. Haque,[3] Natalie Stingelin,[4] Frank C. Spano,[2,*] and Carlos Silva[1,†]

[1] *Département de physique & Regroupement québécois sur les matériaux de pointe, Université de Montréal, C.P. 6128, Succursale centre-ville, Montréal (Québec) H3C 3J7, Canada*

[2] *Department of Chemistry, Temple University, Philadelphia, PA 19122, United States*

[3] *Department of Chemistry and Centre for Plastic Electronics, Imperial College London, South Kensington Campus, London SW7 2AZ, United Kingdom*

[4] *Department of Materials and Centre for Plastic Electronics, Imperial College London, South Kensington Campus, London SW7 2AZ, United Kingdom*

[*] spano@temple.edu
[†] carlos.silva@umontreal.ca; Visiting Professor (Experimental Solid-State Physics), Department of Physics, Imperial College London, South Kensington Campus, London SW7 2AZ, United Kingdom

# Abstract

The electronic properties of macromolecular semiconductor thin films depend profoundly on their solid-state microstructure, which in turn is governed, among other things, by the processing conditions selected and the polymer's chemical nature and molecular weight. Specifically, low-molecular-weight materials form crystalline domains of cofacially π-stacked molecules, while the usually entangled nature of higher molecular-weight polymers leads to microstructures comprised of molecularly ordered crystallites interconnected by amorphous regions. Here, we examine the interplay between extended exciton states delocalized along the polymer backbones and across polymer chains within the π-stack, depending on the structural development with molecular weight. Such two-dimensional excitations can be considered as Frenkel excitons in the limit of weak intersite coupling. We combine optical spectroscopies, thermal probes, and theoretical modeling, focusing on neat poly(3-hexylthiophene) (P3HT) – one of the most extensively studied polymer semiconductors – of weight-average molecular weight ($M_w$) of 3-450 kg/mol. In thin-film structures of high-molecular-weight materials ($M_w > 50$ kg/mol), a balance of intramolecular and intermolecular excitonic coupling results in high exciton coherence lengths along chains (~4 thiophene units), with interchain coherence limited to ~2.5 chains. In contrast, for structures of low-$M_w$ P3HT (<40 kg/mol), the interchain exciton coherence is dominant (~20% higher than in architectures formed by high-molecular-weight materials). In addition, the spatial coherence within the chain is significantly reduced (by nearly 30%). These observations give valuable structural information; they suggest that the macromolecules in aggregated regions of high-molecular-weight P3HT adopt a more planar conformation compared to low-molecular-weight materials. This results in the observed increase in intrachain exciton coherence. In contrast, shorter chains seem to lead to torsionally more disordered architectures. A rigorous, fundamental description of primary photoexcitations in π-conjugated polymers is hence developed: two-dimensional excitons are defined by the chain-length dependent molecular arrangement and interconnectivity of the conjugated macromolecules, leading to interplay between intramolecular and intermolecular spatial coherence.

# I.  INTRODUCTION

'Plastic' semiconductors are often regarded as complex systems as the conjugated macromolecules they are made of can adopt different arrangements and packing, leading to a diverse range of microstructures. These define their electronic properties. For instance, the chain conformation of conjugated polymers and the resulting microstructure has been shown to have a profound impact on the nature of intra- and interchain dispersion of $\pi$-electrons.[1-10] Therefore, it is essential to learn how to manipulate these structural features. One option is to select materials of different molecular weight. From a perspective of classical polymer science, it is well established that polymers of low weight-average molecular weight ($M_w$) form unconnected, extended-chain crystals, usually of a paraffinic-like arrangement.  Due to the non-entangled nature of these relatively short-chain macromolecules, this leads to a polycrystalline, one-phase morphology. In contrast, with high-$M_w$ materials of molecular weight larger than the molecular weight between entanglements ($M_e$), typically two-phase morphologies are obtained, which are comprised of crystalline moieties embedded in largely unordered (amorphous) regions, whereby individual macromolecules bridge multiple domains of order.[11]

Perhaps the central objective and challenge in the broad field of plastic electronics is to relate the electronic properties to the molecular order and thin-film architectures of the constituting materials; hence, significant research efforts have been devoted to that endeavor. Regioregular poly(3-hexylthiophene) (P3HT) has emerged as a model material to investigate the relationship between macromolecular configuration, microstructure and electronic properties. For example, charge transport in P3HT has been found to display clear trends with molecular weight and whether the film displays one-phase (polycrystalline) or two-phase (entangled) morphologies. In field-effect transistors, Kline et al. observed a systematic increase in field-effect mobilities up to a weight-average molecular weight of 40 kg/mol;[12-14] in agreement with reports by Zhang et al.[15] and Koch et al.[16] On the other hand, bulk mobilities display more complex trends with molecular weight, and depend sensitively on film processing routes. Ballantyne et al. found, for instance, an order-of-magnitude decrease in bulk charge mobilities for P3HT of molecular weight $M_w$ >18 kg/mol.[17] Koch et al. reported maximum bulk mobilities for P3HT of $M_w$ ~4 kg/mol. For materials of higher molecular weight, the time-of-flight mobility was found to be again reduced, leveling at a value of $\sim 10^{-4}$ cm$^2$/Vs.[16] However, when films were cast at 115 $^o$C, the mobility *increased* by over an order of magnitude for materials of $M_w$ <10 kg/mol, and *decreased* by over two orders of magnitude for P3HT of $M_w \approx 60$ kg/mol.[16]

In this contribution, we focus on the nature of excitons in regioregular P3HT with molecular weights ranging from 3-450 kg/mol, where the microstructure changes from chain-extended crystals to a semicrystalline morphology where amorphous and crystalline domains are interconnected.[3] This structural transition occurs at a molecular weight of around 25 to 35 kg/mol.[10,18] Our strategy hence is to exploit the intricate influence of electronic structure and spatially correlated energetic disorder on the spectral line shapes in order to extract the microstructure-dependent information on exciton coherence.[19,20] More specifically, our objective is to explore how the exciton coherence lengths, measured along the polymer backbone and across the π-stack, vary with molecular weight and, in particular, with the structural change occurring above the entanglement limit. (The third dimension, which is directed along the lamellar axis, is neglected, as excitonic interactions are negligible between the well-separated π-stacks.) In short, we seek the two-dimensional exciton coherence function as a function of structure.

Conjugated polymers are structurally complex and can, among other things, feature torsional disorder along the backbone, which translates into site energy disorder. Furthermore, the π-electron coupling leading to electronic dispersion is highly anisotropic; dispersion along the chain is present due to intramolecular coupling between adjacent monomer motifs, but interchain coupling also clearly plays a role, leading to two-dimensional dispersion.[21,22] As a consequence, neutral excitations (containing an electron and hole) in this class of materials can be considered to be in the Frenkel limit with respect to interchain interactions, highly localized to a few lattice sites across the lamellar lattice, spanning ~1 nm, and highly influenced by energetic disorder.[19]

Spano has previously described excitons in regioregular P3HT by invoking a weakly coupled H-aggregate model, in which weak excitonic coupling (compared to molecular reorganization energies) between cofacial chromophores in adjacent polymer chains leads to electronic dispersion of the vibronic molecular levels to form vibronic bands, with the bandwidth strongly dependent on microstructure.[23,24] However, the importance of *intra*-chain coupling on optical spectral signatures has become evident by a recent report of J-aggregate like behavior in P3HT nanofibres[1] and in P3HT:PEO complex[25] in which head-to-tail interactions of transition dipole moments of chromophores along the chain influence spectral lineshapes. The limiting case of through-bond, intrachain excitonic coupling leads to what can be considered as Wannier-Mott excitons in one-dimensional lattices.[26,27] In such excitons, the electron and hole are bound over several unit cells. A compelling example of this limit is provided by isolated, extended chains of red-phase polydiacetylene derivatives,[27] which show ultra narrow photoluminescence linewidths and superradiance[28-30] characteristic of J-aggregates at low temperature.[31] The work by Niles et al.[1] therefore demonstrates that in regioregular P3HT films, there is a competition between J-like (intrachain) and H-like (interchain) excitonic coupling.[32] This interplay between through-bond and through-space excitonic coupling of adjacent chromophores must be strongly influenced by energetic disorder and hence the macromolecular conformation and packing in the solid state.

We combine steady-state and time-resolved optical spectroscopy, thermal analysis and theoretical modeling in order to unravel the interplay between through-bond excitonic coupling within polymer chains, and cofacial coupling between adjacent chains within a π-stack. Thus far, the result of such intrachain/interchain exciton coupling has been analyzed theoretically for the case of ordered aggregates with a thermal Boltzmann distribution of emitting excitons.[32] In this work we find that this interplay is strongly dependent on the microstructure imposed by processing and materials parameters such as molecular weight. In the low molecular-weight regime, the paraffinic-like, chain extended structures seems to lead to short chromophores that produce more strongly coupled H aggregates, very likely due to torsional disorder along the polymer backbone introduced, for example, through end-group effects.[11] On the other hand, films prepared with P3HT of molecular weight above the entanglement limit feature weaker interchain coupling but stronger intrachain coupling. We attribute this to the presence of longer-chain chromophore segments originating from torsionally more ordered (i.e. more planar) macromolecules. We thus find that the intrachain and interchain spatial coherence of excitons evolve with opposite trends with molecular weight, with longer intrachain coherence being observed for materials of high $M_w$, but longer interchain coherence in low-molecular weight matter. The photophysical properties arise from the competition between intrachain (through-bond) electronic coupling characteristic of Wannier-Mott excitons, and interchain (through-space) electronic coupling leading to Frenkel excitons, which generally dominate the photophysical properties of plastic semiconductors.

## II. EXPERIMENTAL METHODS

P3HT films of weight-average molecular weight ($M_w$) in the range of 3-450 kg/mol were wire-bar-coated from p-xylene solution (1 wt% solid content) onto glass substrates. The solution temperature was 70 °C, and the substrate was also kept at 70 °C. Absorption spectra were measured using a commercial UV-Vis spectrophotometer (Perkin Elmer, Lambda25). The photoluminescence experiments were carried out by excitation with a continuous-wave laser (Ultralasers Inc., 200 mW maximum, 532 nm) modulated at a frequency of 100 Hz with a mechanical chopper (Terahertz Technologies), and detection with a 300-mm spectrometer (Princeton Instruments SP300 with EOS model S/PBS-025/020-TE2-H photoreceiver), and phase-sensitive detection (SR830 lock-in amplifier). The sample was housed in a closed-cycle, temperature controlled, sample-in-vapor cryostat (CryoIndustries of America).

The ultrafast time-resolved PL measurements were performed by femtosecond fluorescence upconversion. The samples were excited by the frequency-doubled output from an ultrafast mode-locked Ti:Sapphire oscillator (Newport Spectra-Physics Broadband MaiTai). Excitation and gate wavelengths were fixed at 400 and 800 nm, respectively. Both the gate and excitation beams were independently compressed using prism-pair compressors. The excitation intensity was adjusted to be below the onset of intensity dependent kinetics. Photoluminescence from the sample was focused on a 200-µm-thick BBO crystal along with the 800-nm gate beam. Sum frequency generated photons (corresponding to photoluminescence at either 650 or 720 nm) were detected using a photomultiplier tube (R7207-01, Hamamatsu). The temporal resolution of the system was less than 150 fs. Sample degradation was avoided by performing the measurements under flowing nitrogen and using a translation stage to constantly move the sample within the beam, removing the effect of photo-bleaching and providing data averaged across the whole of the sample.

Melting temperatures, $T_m$, and enthalpies of fusion, $\Delta H_f$, were determined with a Mettler DSC822e differential scanning calorimeter (DSC), calibrated with indium and zinc. Samples of 1 - 5 mg (obtained from thin-film structures cast from solution following an identical protocol as for optical measurements) were sealed in aluminum crucibles and then heated under nitrogen at a scanning rate of 10°C/min. Enthalpies of fusion were deduced from the surface area underneath the melting endotherms.

## III. RESULTS AND ANALYSIS

### A.  Relationship between exciton signatures and solid-state microstructure

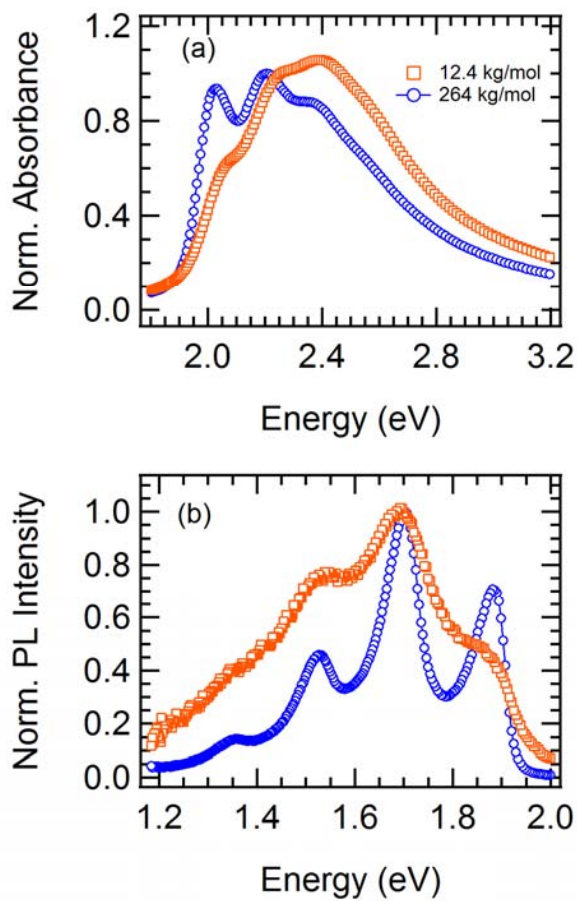

**Fig. 1** Absorption and photoluminescence spectra in thin films of neat P3HT of low (12.4 kg/mol; chain extended structure – red symbols) and high (264 kg/mol; entangled, semicrystalline morphology – blue symbols). (a) Normalized absorption spectra for these films at room temperature. (b) Normalized photoluminescence spectra measured at T = 10 K.

In this section, we explore the dependence of steady-state absorption and photoluminescence (PL) spectral line shapes of P3HT films with $M_w$. Fig. 1(a) displays the room-temperature absorption spectra of films of P3HT of two different $M_w$. Here we show spectra of two films, one prepared with P3HT of low $M_w$ (12.4 kg/mol) and one made of high-$M_w$ material (264 kg/mol). We discern two distinct differences in the spectral line shapes of these samples. The first concerns the relative absorbance of the origin of the vibronic progression at 2.0 eV. As mentioned in Section I, previous work by Spano has demonstrated that the absorption lineshape can be understood within the framework of a weakly coupled H-aggregate model.[5,6,23] Weak resonant Coulombic coupling between cofacial polymer chains, $J_{inter}$, results in electronic dispersion of the vibronic molecular levels to form vibronic bands with free-exciton bandwidth $W = 4J_{inter}$. $W$ is related to absorbance ratio of 0-0 and 0-1 peaks at 2.0 and 2.2 eV, respectively. Assuming a Huang-Rhys parameter of unity, the ratio is given by[5,6,23]

$$\frac{A_{0-0}}{A_{0-1}} \approx \left( \frac{1 - 0.24 W / \hbar\omega_0}{1 + 0.073 W / \hbar\omega_0} \right)^2 \quad , \tag{1}$$

where $\hbar\omega_0 = 180$ meV is the effective energy of the main intramolecular vibrational modes coupled to the electronic transition. Our optical absorption data presented in Fig. 1(a) thus suggests that $W$ decreases significantly in high-$M_w$ materials compared to P3HT of low $M_w$. We associate this noticeable variation in $W$ between the two samples to conformational changes resulting from the significantly different average chain lengths associated with these two films. Excitonic coupling has been reported to vary inversely with conjugation length in P3HT,[4,33] consistent with theoretical predictions.[34] Similar trends have been reported in a series of chiral oligothiophenes end-capped with enthylene oxide side chains[33] and by quantum-chemical calculations.[34]

The second important difference in the absorbance spectra of low- and high-$M_w$ materials is the change in relative spectral weight at photon energies around 2.4 eV, which is higher for the sample of lower $M_w$. Absorbance in this high-energy region can be largely assigned to electronically uncoupled chromophores,[6,19] implying that the lower $M_w$ sample is comprised a higher fraction of photophysically uncoupled polymer chains.

Further information can be obtained from the corresponding PL spectra of the two low- and high- $M_w$ samples, measured at 10 K (Fig. 1(b)). Specifically, we observe a 0-0 (1.88 eV)/0-1 (1.70 eV) PL ratio that is significantly larger in the P3HT film of higher $M_w$; we discuss this trend in more detail later in this section. In addition, we note that the 0-2 (1.52 eV)/0-1 (1.70 eV) intensity ratio also depends on $M_w$; the smaller ratio for the higher $M_w$ samples indicating a *smaller* Huang-Rhys parameter. This is somewhat surprising as it is inconsistent with the weakly-coupled H-aggregate model,[19] which predicts that in the weak excitonic coupling regime the *relative* intensities of the 0-1, 0-2,… satellites should be independent of $W$ (and disorder) and, hence, should resemble what is obtained for an isolated chromophore. Aggregation mainly influences the relative 0-0 peak intensity which depends entirely on the coherence of the emitting exciton, in marked contrast to the mainly incoherent origin of the vibronic satellites.

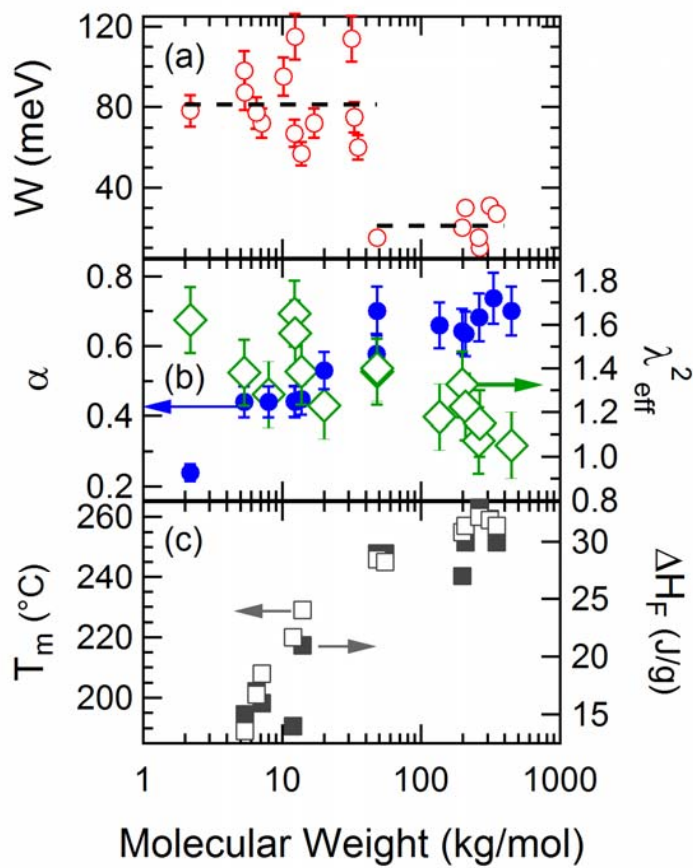

**Fig.2** Dependence of spectroscopic and microstructural features and parameters from spectral analysis on molecular weight of neat P3HT architectures. (a) Free-exciton bandwidth, $W$, derived from absorption spectra at room temperature. (b) Relative intensity of the 0-0 PL band, $\alpha$, and effective Huang-Rhys parameter, $\lambda_{eff}^2$, derived from photoluminescence spectra at 10 K. (c) Melting temperature, $T_m$, and enthalpy of fusion, $\Delta H_f$, derived from differential scanning calorimetry measurements.

In order to examine these trends more quantitatively we measured absorption and PL spectra in a series of neat P3HT thin films using materials of $M_w$ ranging from 2 to 347 kg/mol. We first plot $W$ versus $M_w$ (Fig. 2(a)), derived from eq. 1. For materials of $M_w < 50$ kg/mol, we deduce an average $W = 81 \pm 11$ meV, while for P3HT of higher $M_w$ we consistently derive an average $W$ of $21 \pm 2$ meV. Tellingly, this transition of $W$ is abrupt and occurs around $M_w \approx 50$ kg/mol – which is the molecular-weight region where the P3HT chains attain sufficient length to undergo entanglement, leading to a microstructural change from a paraffinic, chain extended phase to a two-phase morphology composed of interconnected crystalline and amorphous regions.[10,12,16,18,35] From the PL spectra of all the samples, we obtain the 0-0/0-1 intensity ratio and the *effective* Huang-Rhys parameter as a function of $M_w$ via a fit with modified Franck-Condon model (see below) that takes into account the H-aggregate nature of the PL spectrum, namely, a symmetry-forbidden but disorder-allowed intensity of the 0-0 peak at 1.88 eV:[6]

$$I(\omega) \propto (\hbar\omega)^3 n^3(\omega) e^{-\lambda_{\text{eff}}^2} \times \left[ \alpha\Gamma(\hbar\omega - E_0) + \sum_{m=1,2,...} \frac{\lambda_{\text{eff}}^{2m}}{m!} \Gamma(\hbar\omega - E_0 - m\hbar\omega_0) \right], \qquad (2)$$

where $n(\omega)$ is the refractive index of the film at the optical frequency $\omega$, $\lambda_{\text{eff}}^2$ is the effective Huang-Rhys parameter, $E_0$ is the energy of the origin of the vibronic progression, $\hbar\omega_0 = 180\,\text{meV}$ is the energy of the effective oscillator coupled to the electronic transition, and $\Gamma$ is a Gaussian function that represents the inhomogeneously broadened spectral line of the vibronic replica in the progression. The parameter $\alpha$ thus quantifies the relative intensity of the origin of the vibronic progression (0-0), which is decoupled from the rest of the progression because it alone expresses the spatial coherence of the emitting Frenkel exciton within the H-aggregate model.[19] We display $\alpha$ and $\lambda_{\text{eff}}^2$ as a function of $M_w$ in Fig. 2(b). As the $M_w$ of the P3HT films increases, we observe a strong increase in $\alpha$ from 0.24 to 0.74, correlated with a decrease of $\lambda_{\text{eff}}^2$ from ~1.5 to just over 1 for high-$M_w$ samples. Note that for the analysis of the absorption spectrum reported in Fig. 2(a), we assumed $\lambda_{\text{eff}}^2 = 1$. We also extracted values for $W$ using $\lambda_{\text{eff}}^2$ from the PL analysis in Fig. 1(b), using a more general expression reported elsewhere[24] (see supp. information), generalized for arbitrary HR factors. This increases $W$ by ~50%. However, we expect that $\lambda_{\text{eff}}^2$ in the absorption spectrum is intermediate between 1 and that measured by PL since the distribution of chromophore configurations is larger in absorption. Therefore, we choose to report $W$ in Fig. 2(a) obtained by imposing $\lambda_{\text{eff}}^2 = 1$.

The increase in $\alpha$ – the relative 0-0 intensity - can be accounted for in the H-aggregate model as a decrease in $W$, an increase in the energetic disorder magnitude (quantified by the width $\sigma$ of a Gaussian distribution of site energies), a decrease in spatial correlation of site energies ($\beta$), or some combination of the above. Here, $\beta \equiv \exp(-1/l_0)$ where $l_0$ is the site-energy spatial correlation length in units of the interchain separation. (Note that $\beta$ ranges from 0 to 1 as $l_0$ increases from 0 to infinity.) From the perturbative expression derived in Ref. [23] we have

$$\frac{I^{0-0}}{I^{0-1}} \sim \frac{(1-\beta)\sigma^2}{(1+\beta)W^2}, \qquad (3)$$

which is strictly valid in the limit where $\sigma << W << \hbar\omega_0$. Since the absorption and emission spectral linewidths are smaller in the higher-$M_w$ films, the width of the disorder distribution is also smaller. A change in $\sigma$ therefore cannot be responsible for the higher 0-0/0-1 ratio observed in the high $M_w$ materials. Moreover, increasing order is usually coincident with an increasing spatial correlation parameter $\beta$, which also serves to decrease the 0-0/0-1 ratio and hence contradicts what we observe.

A decrease in the spatial correlation parameter $\beta$ could contribute to the trend observed in the spectroscopic parameters displayed in Fig. 2(b), and we will explore this in greater detail later in section IIIB. (We find, in fact, that $\beta$ increases for films prepared with higher $M_w$ P3HT.) The data displayed in Fig. 2(a) shows, however, that $W$ is significantly smaller for higher-$M_w$ samples (consistent also with increasing order) and thus, according to the H-aggregate model, a reduction in $W$ would account for the observed increase in the PL 0-0/0-1 ratio with $M_w$. Because of the large change in $W$ in the two well-defined $M_w$ regimes, we expect that this parameter will represent the dominant contribution to the trend in $\alpha$. In the limit that $W$ becomes negligible (i.e. in $\pi$-stacks consisting of polymers with very long conjugation lengths), the 0-0/0-1 line strength ratio is capped at the value $\lambda^2$, the Huang-Rhys parameter representing a single chain in the $\pi$-stack. Measurements of Kanemoto et. al.[36] show that the single chain HR factor is approximately unity.

There is, however, evidence for the breakdown of the pure H-aggregate model in the recent literature. For instance, recent reports on P3HT whiskers (i.e. 'nanofibers' grown from relatively high molecular weight materials from solution over time), have shown that the ratio of the 0-0 to 0-1 PL peaks can be as high as two at room temperature,[1] suggesting that the line strength can significantly exceed the aforementioned cap. This cannot be accounted for in the H-aggregate model. Additional evidence for the breakdown of the H-aggregate model comes from the decrease of the 0-2/0-1 peak ratio in the PL spectrum of materials of increasing $M_w$ reported in this paper (Fig. 1(b)). As mentioned in a prior work,[19] for the relatively weak values of $W$ encountered in P3HT aggregates the *relative* strengths of the side-band peaks (0-1, 0-2, etc.) in the PL spectrum very closely resemble the isolated molecular values; for example, the ratio of the 0-2 to 0-1 line strengths, $I^{0-2}/I^{0-1}$, remains steadfast at the value of $\lambda^2/2$, as determined from the Poissonian distribution for an isolated (single molecule) chain:

$$I_{molec}^{0-n} = \frac{\lambda^{2n}e^{-\lambda^2}}{n!}. \qquad (4)$$

This relation is in disagreement with the data that is presented in Figs. 1(b) and 2(b), where we observe a significant decrease in the PL 0-2/0-1 ratio (decrease in $\lambda_{eff}^2$) with increasing $M_w$. More precisely, $\lambda_{eff}^2$ decreases from approximately 1.3-1.6 for P3HT of low $M_w$ to 1.0-1.1 for materials of high $M_w$. The apparent failure of the H-aggregate model in this respect is due to its assumption that a single chain can be modeled as a two-level chromophore with displaced harmonic nuclear potentials characterized by an HR factor $\lambda^2$. This simple description assumes that the Born-Oppenheimer approximation is valid for a single chain and ignores exciton motion and nonadiabatic coupling to vibrations within a given chain. The consequences of *intrachain* exciton vibrational coupling were examined in Ref.[31], where a single conjugated polymer chain was modeled as a one-dimensional Wannier exciton coupled vibronically to the main symmetric vinyl-stretching mode. The electron and hole transfer integrals between neighboring unit cells were taken to have the same sign, consistent with what is normally assumed for conjugated polymer chains, and reflective of a direct bandgap semiconductor. The ensuing photophysical properties of such 'quantum wires' were found to be remarkably similar to linear J-aggregates.[37] The model quantitatively accounted for all of the salient photophysical features exhibited by isolated chains of red-phase polydiacetylene, including the very large 0-0/1-0 absorption and PL spectral ratios, as well as the $1/\sqrt{T}$ dependence of the latter. In Ref.[31] it was also shown that the 0-0/0-1 line strength ratio in the PL spectrum of the single chain increases with intrachain exciton bandwidth — as expected for a J-aggregate — while the 0-2/0-1 ratio decreases. The behavior can be crudely described as a decrease in the effective single chain Huang-Rhys factor with increasing exciton coupling, although the relative intensities are generally not Poissonian. An approximate description of this behavior was first provided by Chen and coworkers for a Frenkel chain.[38] We therefore attribute the decrease in $\lambda_{eff}^2$ that we observe in P3HT sample of increasing $M_w$ to an *increase* in the intrachain exciton bandwidth. This strongly suggests that in these high-$M_w$ P3HT architectures, the macromolecules are of reduced torsional disorder. Such a transition to a more planar structure in samples of higher molecular weight would be agreement with some of our thermal analysis data: indeed, we observe an increase in both the melting temperature, $T_m$ (which is directly correlated with the lamellar crystal thickness) and the enthalpy of fusion, $\Delta H_f$ (which is directly correlated with the degree of crystallinity; see Koch et al.[16] and references therein).

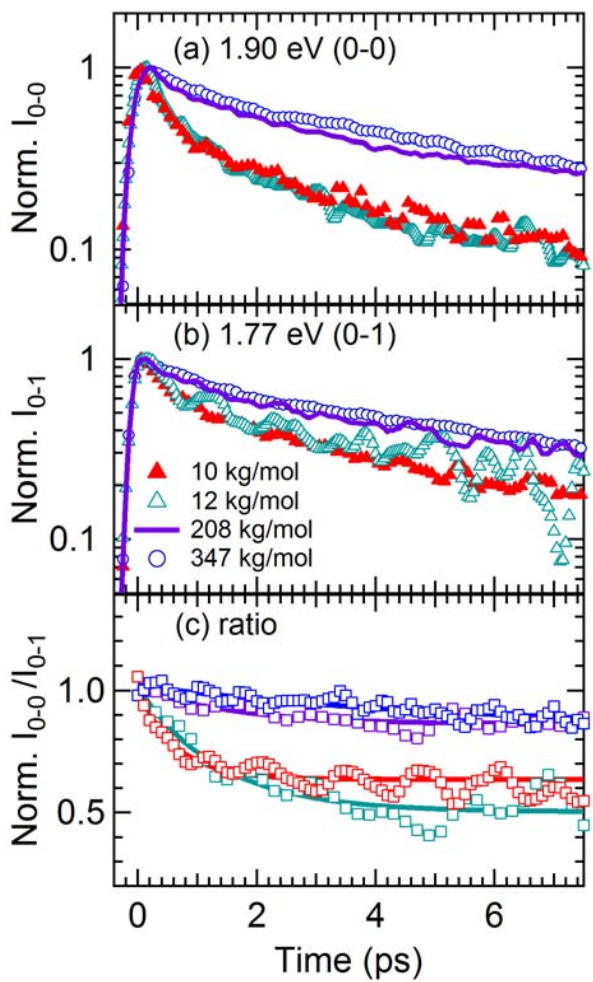

**Fig.3** (Color) Time-resolved photoluminescence intensity measured at room temperature for various neat P3HT films, measured at photon energy of 1.90 eV (a) and 1.77 eV (b). The normalized ratio $I^{0-0} / I^{0-1}$ is plotted in part (c).

Further support for P3HT structures of high-$M_w$ P3HT structures being of higher torsional order (i.e more pronounced planarization of the polymer backbones) can be evoked from the ratio of the PL decay at 1.90 eV and 1.77 eV (Fig. 3c). We observe that in the low $M_w$-regime, the relaxation of the normalized ratio of intensities at these two photon energies is more significant than for high-$M_w$ material. In fact, Parkinson et al. have reported that vibrational relaxation from a low-symmetry to a high-symmetry ordered state, characteristic of torsional planarization along the polymer backbone upon optical excitation, results in a dynamic decrease of the relative 0-0 PL intensity on a timescale ~13ps.[39] Such effect has also been observed by Banerji et al. on timescales >1 ps.[40] Moreover, recent broadband transient spectroscopy measurements have revealed signatures of both small-scale (<1 ps) and large scale (>1 ps) planarization of P3HT thiophene units after excitation.[41] In other materials such as PPV oligomers, the degree of torsional angle of the backbone is found to correlate with an increase of the oscillator strength for the 0-0 transition in comparison to the rest of the vibronic replicas.[42] This is consistent with higher initial torsional order in the high $M_w$ regime, which influences $W$ and consequently $\alpha$, the relative intensity of the 0-0 peak in PL. From this observation, we support our conjecture that the decrease in $\lambda_{eff}^2$ with increasing $M_w$ is primarily due to an increase in intrachain exciton bandwidth resulting from a reduction in torsional disorder in high $M_w$ aggregates.

## B. Relationship between exciton signatures and solid-state microstructure

In order to obtain a more detailed knowledge about the relationship between exciton signatures and solid-state microstructure and macromolecular chain conformation, and to explore the effect of intrachain exciton bandwidth on $\lambda_{eff}^2$ quantitatively, we introduce a variation on the HJ-aggregate model introduced in Ref.[32] In the HJ-aggregate model, excitons are free to navigate across chains in the $\pi$-stack, as in the original H-aggregate model, but also *within* chains. Intrachain excitons are generally of the Wannier-Mott type in which electrons and holes are bound across more than one site with a characteristic Bohr radius. Hence, within a $\pi$-stack of such chains, the exciton is delocalized over two dimensions; along the polymer backbone as well as along the stacking axis. Because the Wannier-Mott excitons in one dimension lead to J-aggregate photophysics,[31] the aggregate can be understood as J-like along the backbone dimension and H-like along the stacking direction, allowing for interesting hybrid H-J photophysical properties. If disorder is absent, as was the case treated in Ref.[32], the high symmetry dictates that no 0-0 emission can occur at low temperatures – a classic signature of H-aggregates. Increasing temperature leads to interesting effects, for example, a transition from H- to J-aggregate behavior with increasing temperature has been predicted.[32] Similar effects are expected when correlated disorder (quantified by $\sigma$ and $\beta$) is included.

In conjugated polymer chains the ratio of the 0-0 and 1-0 oscillator strengths in the absorption spectrum can be quite large – almost a factor of ten in the isolated and practically disorder-free polydiacetylene (PDA) chains,[30] for example – and is a direct measure of the strength of the intrachain exciton coupling. In modeling a single P3HT chain we assume weaker intrachain coupling than in PDA because the 0-0/1-0 absorption peak ratio in P3HT (in solution or solid phase) has never been observed to increase much beyond unity, in marked contrast to the case of PDA. In the weak coupling limit, second order perturbation theory shows that transport along the chain occurs via virtual two-step electron-hole (or vice versa) events involving Frenkel-like excitations within a single monomer unit (i.e. thiophene unit).[32] Such a chain can be described with a Frenkel-like Holstein Hamiltonian in which the (through-bond) exciton coupling between adjacent thiophene units is given by

$$J_{intra} = -\frac{2t_e t_h}{U - V_1}.$$ (5)

Eq. (5) contains the product of electron ($t_e$) and hole ($t_h$) transfer integrals connecting neighboring thiophene LUMO's and HOMO's respectively. $U - V_1$ is the stabilization of the neutral Frenkel-like exciton relative to the nearest-neighbor charge-separated state. Note that the single-chain physics is invariant to a sign change in *both* $t_e$ and $t_h$. In addition to an effective coupling between adjacent thiophene units there is also a second-order energy (red) shift, given by

$$\Delta_{intra} = -\frac{2(t_e^2 + t_h^2)}{U - V_1}$$ (6)

due to virtual two-step transfer where, for example, an electron (or hole) moves to a neighbor and then returns to the parent thiophene ring.

As in Ref. [32], we assume equivalent electron and hole transfer integrals, and introduce $t$ such that $t \equiv t_e = t_h$. Such an approximation is consistent with the direct bandgap nature of conjugated polymer chains. Within this approximation

$$J_{intra} = -2t^2 / (U - V_1),$$  (7a)

and the red-shift in Eq.(6) reduces to

$$\Delta_{intra} = 2J_{intra}.$$  (7b)

Since the P3HT chains are generally torsionally disordered, we consider $t$ (and $J_{intra}$) to be an *effective* coupling, smaller than the larger intrinsic coupling found in the "perfect" polydiacetylene chains.[31] The negative sign on the right-hand side of eq. 7a ensures J-aggregate-like behavior.

Within the space of a single electronic excitation (one electron and one hole), the single-chain effective Hamiltonian representing the $s$th disorder-free P3HT chain is therefore given by

$$\begin{aligned}
\hat{H}_{s,chain} = \varepsilon_1 + \Delta_{intra} + \hbar\omega_0 \sum_n b_{n,s}^\dagger b_{n,s} + \hbar\omega_0 \sum_n \left\{ \lambda_0 (b_{n,s}^\dagger + b_{n,s}) + \lambda_0^2 \right\} |n,s\rangle\langle n,s| \\
+ J_{intra} \sum_n \left\{ |n,s\rangle\langle n+1,s| + |n+1,s\rangle\langle n,s| \right\}
\end{aligned};$$  (8)

$|n,s\rangle$ represents an electron-hole (Frenkel-like) excitation ("$S_1$") on the $n$th thiophene unit of the $s$th chain with energy $\varepsilon_1$. The operator $b_{n,s}^\dagger$ ($b_{n,s}$) creates (destroys) a vibrational quantum of energy, $\hbar\omega_0$, in the ground state nuclear potential well corresponding to the absence of electrons and holes in repeat unit $n$ (the "$S_0$" potential). The quantity $\lambda_0^2$ is the Huang-Rhys factor for an *individual* thiophene unit and is set to 2, which we calculate for uncoupled thiophene monomers by means of time-dependent density functional theory (see Supplementary Information).

Energetic site disorder within the $s^{\text{th}}$ chain is represented by the Hamiltonian

$$H_{s,dis} = \sum_n \Delta_{n,s} |n,s\rangle\langle n,s| \quad , \qquad (9)$$

where $\Delta_{n,s}$ is the transition frequency detuning of the $n^{\text{th}}$ thiophene unit on the $s^{\text{th}}$ chain. In what follows, we take the detunings, $\Delta_{n,s}$, to be Gaussian random variables with spatial correlation assumed to be isotropic across the two dimensions defining the π-stack. The Gaussian width (full width at $1/e$) is $2\sigma$ and the spatial correlation length is $l_0$. The latter is in units of $d$, where $d \approx 4\,\text{Å}$ is the (approximate) distance between nearest-neighbor chains as well as the distance between nearest neighbor thiophene units within a given chain. The monomer units in a π-stack therefore constitute a square lattice so that $l_0$ is more properly defined as the correlation *radius*. Given $l_0$, the correlation parameter $\beta$ follows from $\beta = \exp(-1/l_0)$. In a more sophisticated treatment, the interchain and intrachain disorder included in Eq. 9 can be distinguished by employing separate disorder correlation lengths along and across the polymer backbone.

The complete Hamiltonian for the π-stack including disorder reads,

$$H_{\pi-stack} = \sum_s (\hat{H}_{s,chain} + H_{s,dis}) + \sum_s \sum_n J_{inter} \left\{ |n,s\rangle\langle n,s+1| + |n,s+1\rangle\langle n,s| \right\} \quad , \qquad (10)$$

where the through-space, interchain coupling, $J_{inter}$, is limited to adjacent (same $n$) thiophene units on neighboring chains. Note that torsional disorder is also reflected in the *mean* intrachain coupling $J_{intra}$. In principle $J_{intra}$ should be correlated to $l_0$. Here we treat them as independent variables but look for consistency in our final results (see below). Note that the H-aggregate model is recovered in the limit that $J_{intra}$ is set to zero.

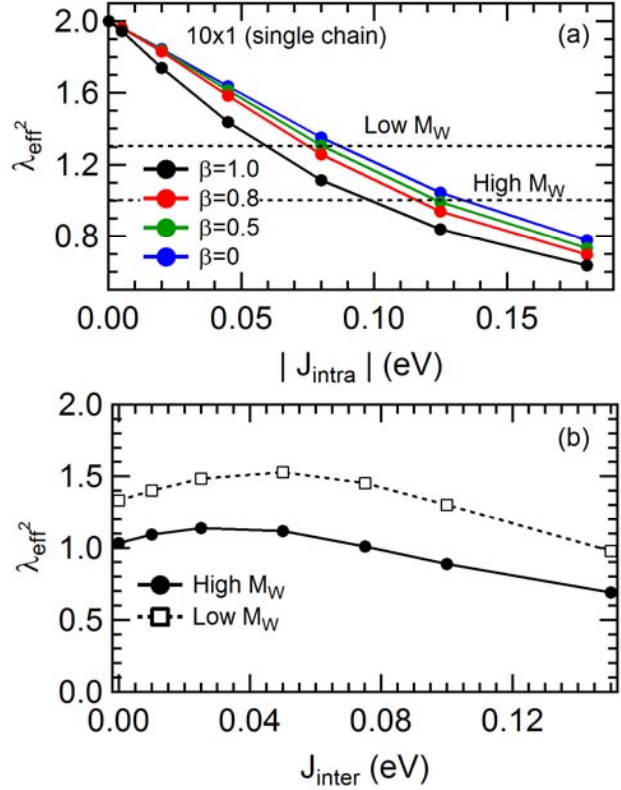

**Fig. 4** (Color) Calculated effective Huang-Rhys parameter as a function of intramolecular and intermolecular coupling. (a) $\lambda_{eff}^2$ as a function of $\mid J_{intra} \mid$ for a single polymer chain containing 10 thiophene units with $\lambda_{eff}^2 = 2$. Several disorder parameters $\beta$ are shown, with $\sigma = 0.54 \ \hbar\omega_0$. (b) $\lambda_{eff}^2$ as a function of interchain coupling for 10 by 6 aggregates of high and low molecular weight chains with β=0.6.

In order to derive the mean intramolecular coupling $J_{intra}$ from the measured PL spectra we first introduce the effective HR factor, defined as,

$$\lambda_{eff}^2 \equiv 2 \frac{I_{PL}^{0-2}}{I_{PL}^{0-1}},$$ (11)

where $I_{PL}^{0-v}$ are the line strengths corresponding to the 0-$v$ transition. The evaluation of the line strengths from a thermal population of exciton emitters is outlined in Ref. [32]. The effective HR factor defined by Eq. 11 is a more direct measure of the intrachain exciton bandwidth, similar to the $A_{0-0}/A_{1-0}$ ratio in the absorption spectrum. The advantage of using $\lambda_{eff}^2$ in Eq. 11 is that it is a characteristic of the *emitting* species and it is quite likely that the absorbing and emitting species are different in P3HT films, where disorder enhances spectral diffusion away from the primary absorbers. The effective HR factor can also be defined as the ratio, $I_{PL}^{0-1}/I_{PL}^{0-0}$. However, this ratio is sensitive to the exciton coherence number, $N_{coh}$, which may not be a direct measure of exciton bandwidth. For example, in the limiting case of a disorder-free H-aggregate at T = 0 K, $I_{PL}^{0-0} = 0$ by symmetry, *independent* of the exciton bandwidth. The 0-0/0-1 ratio is very sensitive to the destructive interference between polymer chains in the p-stack, thereby obscuring the dependence on the intrachain bandwidth.

In the limit when $J_{intra} = 0$, a single chain becomes a collection of noninteracting thiophene units and $\lambda_{eff}^2$ in Eq. 11 reduces to $\lambda_0^2$ (which we take to be two), as can be appreciated from the Poissonian distribution in Eq. 2. Increasing $|J_{intra}|$ in a *single* chain enhances the J-aggregate behavior: the 0-0 PL intensity increases, while the 0-2 intensity *decreases*, both relative to the 0-1 intensity. The overall result is a decrease in the effective HR factor. The effect is shown in Fig. 4 for several values of the disorder correlation parameter, $\beta$, with the value of $\sigma$ taken from the absorption spectral line widths of the high-$M_w$ films when fitted to a Gaussian line shape. ($\sigma$ is the half-width at 1/$e$ of the maximum). From Fig. 1a we have $\sigma = 0.6\omega_0$ and $0.54\omega_0$, for the low and high $M_w$ samples, respectively. Fig. 4 shows that $\lambda_{eff}^2$ decreases markedly with increasing intrachain coupling, as expected for J-aggregates.

After removing the cubic frequency dependence and the index of refraction dependence from the measured PL spectra in Fig. 1b, the values of $\lambda_{eff}^2$ were determined to be approximately 1.5 for the low $M_w$ materials and 1.1 for P3HT of high $M_w$. However, in our modeling of a single chain, we take the slightly smaller values of 1.3 and 1.0, for the low- and high-$M_w$ films, respectively, since, as will be shown shortly, the values increase somewhat when interchain effects are included. Fig. 4 shows that $\lambda_{eff}^2$ attains the value of near unity observed in films of P3HT of high $M_w$ when $|J_{intra}|$ is in the range, 0.1-0.15 eV, while the value observed for low-$M_w$ material is consistent with values of $|J_{intra}|$, some 30% smaller.

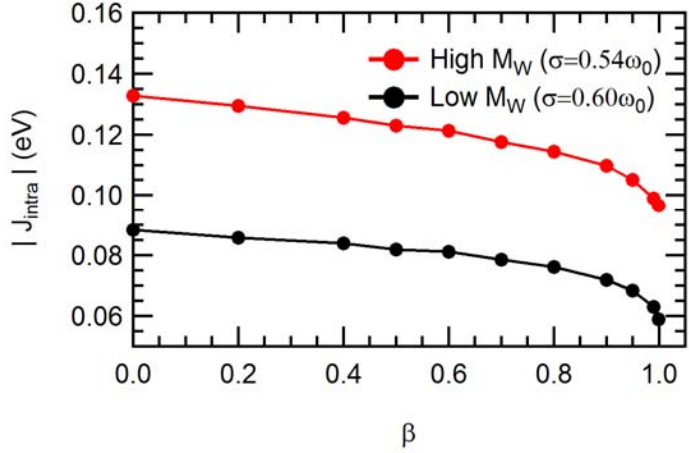

**Fig.5** Values of ($\beta$, $J_{intra}$) for which the effective HR factor, $\lambda_{eff}^2$ is constant at 1.0 and 1.3 for single (isolated) chains of low and high $M$w, respectively. Each chain contains 10 thiophene rings (sufficient for convergence) and the ensemble average included $10^4$ configurations. For the low (high) $M$w chains the value of $\sigma$ is $0.60\omega_0$ ($0.54\,\omega_0$) as determined from the absorption spectral line widths.

Fig. 4(b) shows how $\lambda_{eff}^2$ varies with *inter*chain coupling, $J_{inter}$, for the two values of $|J_{intra}|$ which reproduce the measured effective HR factors for low and high $M_w$-materials when $\beta = 0.6$. The dependence on $J_{inter}$ is weaker than the dependence on $|J_{intra}|$, although there is a slight but significant initial rise in $\lambda_{eff}^2$ with $J_{inter}$, followed by a peak and subsequent demise. The weak dependence of $\lambda_{eff}^2$ on interchain interactions was shown analytically using the H-aggregate model in Ref.[23] and demonstrated numerically in Ref.[19] The weak dependence is therefore due to the positive sign of the interchain coupling.

Fig. 4 (a) and (b) also show that $\lambda_{eff}^2$ serves as an effective probe of the *intra*chain exciton bandwidth and disorder parameter $\beta$. Armed with the measured value of $\lambda_{eff}^2$ for the low and high $M_w$ films (1.3 and 1.0, respectively) we then determined the points ($\beta$, $J_{intra}$) which reproduce the measured $\lambda_{eff}^2$ in an ensemble of disordered *single* chains containing 10 thiophene units each. (Ten units are enough to ensure convergence to the polymer limit.) The results are plotted in Fig. 5 for both low- and high-$M_w$ films. Our results show that generally slightly larger values of $J_{intra}$ are required when disorder is increased ($\beta$ reduced) and that the low-$M_w$ films have uniformly lower values of $|J_{intra}|$ than films prepared with high-$M_w$ materials, consistent with increased torsional disorder in the low-$M_w$ films.

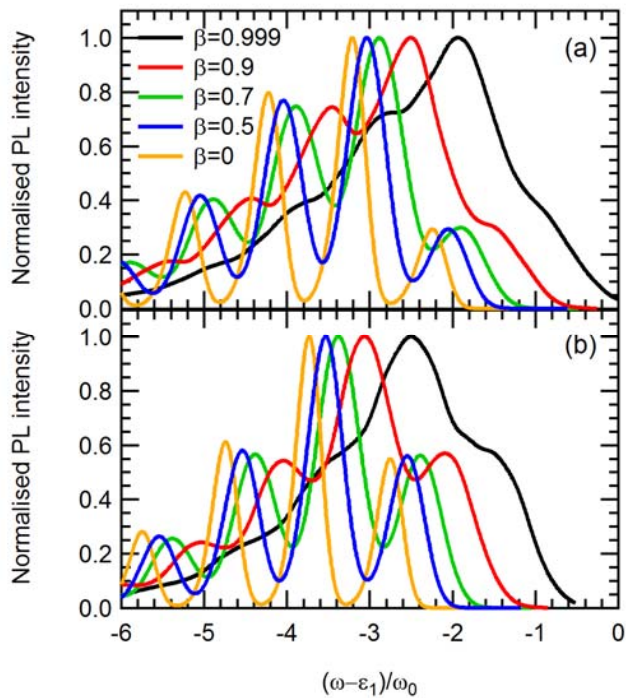

**Fig. 6** Calculated PL spectra for selected triple points from Table 1 for materials of molecular weights in the low (a) and high $M_w$ (b) limits. Ensemble averaging includes $10^4$ configurations of disorder.

We next turn to the effects of interchain coupling. To this end, we first took representative points on the curves in Fig. 5 and varied $J_{inter}$ on each one until the calculated 0-0/0-1 PL ratio matched the measured values. Unlike $\lambda_{eff}^2$, the 0-0/0-1 PL ratio is extremely sensitive to exciton spatial coherence[19,20] and depends in a complex way on the exciton coupling along and across chains.[32] For the two featured P3HTs of low (12.4 kg/mol) and high (264 kg/mol) $M_w$, the 0-0/0-1 ratios were approximately 0.3 and 0.55, respectively, after removal of the cubic frequency dependence and index of refraction dependence. The values of $J_{inter}$ which maintain the measured 0-0/0-1 PL ratio are shown in Table 1. Ideally, if interchain interactions had no impact on $\lambda_{eff}^2$, all of the "triple points" ($\beta$, $J_{intra}$, $J_{inter}$) in Table 1 would reproduce both the measured 0-2/0-1 and 0-0/0-1 ratios in the PL spectrum. Table 1 shows that $\lambda_{eff}^2$ increases with the disorder as anticipated from Fig. 2. The increase, however, is quite modest and within the experimental error bars for both (low- and high-$M_w$ samples) (1.5 and 1.1) when $\beta > 0.5$.

**Table 1: Values of β and |$J_{intra}$| from Fig. 5 along with values of $J_{inter}$ which maintain the 0-0/0-1 PL ratio at 0.56 in films of high $M_w$ P3HT, and 0.30 in low-$M_w$ films.**

| | | High $M_w$ | | | | Low $M_w$ | | | |
|---|---|---|---|---|---|---|---|---|---|
| $\beta$ | $l_0$ | $|J_{intra}|$(eV) | $J_{inter}$(eV) | $I^{0-0}/I^{0-1}$ | $\lambda^2_{eff}$ | $|J_{intra}|$(eV) | $J_{inter}$(eV) | $I^{0-0}/I^{0-1}$ | $\lambda^2_{eff}$ |
| 0 | 0.0 | 0.133 | 0.045 | 0.56 | 1.24 | 0.088 | 0.085 | 0.3 | 1.62 |
| 0.2 | 0.6 | 0.129 | 0.041 | 0.56 | 1.21 | 0.086 | 0.078 | 0.3 | 1.60 |
| 0.4 | 1.1 | 0.126 | 0.035 | 0.56 | 1.18 | 0.084 | 0.069 | 0.3 | 1.56 |
| 0.5 | 1.4 | 0.123 | 0.032 | 0.56 | 1.17 | 0.082 | 0.065 | 0.3 | 1.54 |
| 0.6 | 2.0 | 0.121 | 0.030 | 0.56 | 1.14 | 0.081 | 0.061 | 0.3 | 1.51 |
| 0.7 | 2.8 | 0.117 | 0.028 | 0.56 | 1.13 | 0.078 | 0.055 | 0.3 | 1.50 |
| 0.8 | 4.5 | 0.114 | 0.024 | 0.56 | 1.10 | 0.076 | 0.048 | 0.3 | 1.46 |
| 0.9 | 9.5 | 0.110 | 0.020 | 0.56 | 1.07 | 0.072 | 0.039 | 0.3 | 1.42 |
| 0.95 | 19.5 | 0.105 | 0.015 | 0.56 | 1.05 | 0.068 | 0.030 | 0.3 | 1.39 |
| 0.99 | 99.5 | 0.099 | 0.008 | 0.55 | 1.03 | 0.063 | 0.016 | 0.3 | 1.34 |
| 0.999 | 999.5 | 0.097 | 0.003 | 0.57 | 1.01 | 0.058 | 0.006 | 0.3 | 1.32 |

The Table also shows that as disorder increases along with |$J_{intra}$|, the value of $J_{inter}$ required to keep the 0-0/0-1 ratio constant also increases. This follows because the 0-0/0-1 ratio embodies a competition between J-favoring interactions (|$J_{intra}$|) and H-favoring interactions ($J_{inter}$)[32]; hence, increasing |$J_{intra}$| *increases* the 0-0/0-1 ratio as in J-aggregates, while increasing $J_{inter}$ *decreases* the 0-0/0-1 ratio as in H-aggregates.[20]

Calculated PL spectra for selected triple points are shown in Fig. 6. The 0-0/0-1 ratio is constant throughout the series, whereas the 0-2/0-1 ratio exhibits the aforementioned slight rise as disorder increases. The line widths significantly red-shift and narrow with increasing disorder due to a statistical effect discussed in detail in Ref.[19]. Essentially, the probability of finding a low-energy emitting trap increases when the spatial correlation length $l_0$ is smaller ($\beta$ smaller). Comparison with the experimental linewidths for films of low- and high- $M_w$ P3HT(see Fig. 1) suggests that $\beta$ is in the range, $0.5 < \beta < 0.9$, equivalent to a correlation length in the range. $1 < l_0 < 10$. The measured spectra of Fig. 1(b) further show that the 0-0 transition energies of the low- and high- $M_w$ samples are approximately the same. To obtain specially aligned 0-0 peaks in the PL spectra in Fig. 4 requires $\beta$ to be larger in the higher $M_w$ samples; for example, when $\beta = 0.9$ in the high $M_w$ films and $\beta = 0.5$ in the low-$M_w$ films the positions of the 0-0 peaks are roughly the same. This is consistent with films of P3HT of low $M_w$ featuring a higher degree of energetic disorder.

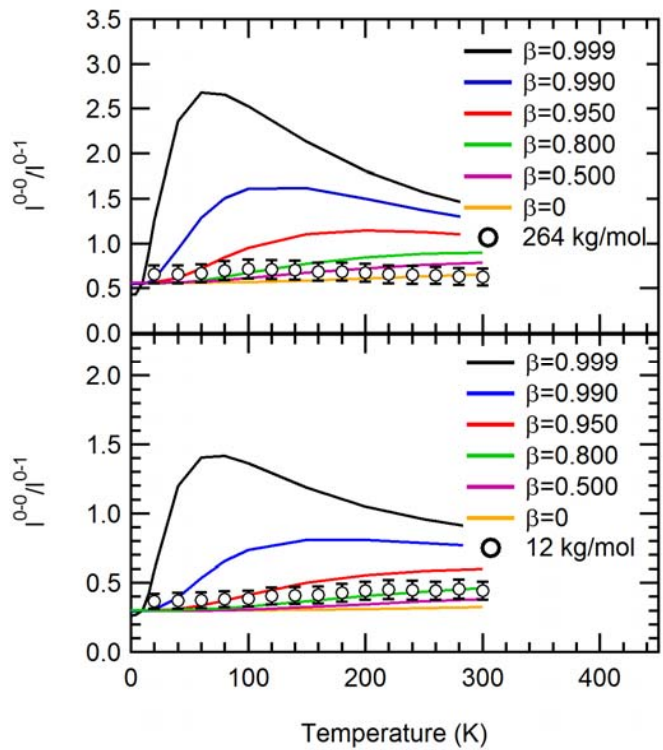

**Fig.7** The 0-0/0-1 PL ratio as a function of temperature for selected triple points (see Table 1) for P3HT thin films of high and low $M_w$. Also shown is the experimental data for the 447 kg/mol film (a) and 8 kg/mol (b) at T=10K.

We now turn to the temperature dependence of the 0-0/0-1 ratio. The measured temperature dependence is shown in Fig. 7 for P3HT of low (12 kg/mol) and high (264 kg/mol) $M_w$, alongside the calculated temperature dependence for the selected triple points from Table 1. The temperature dependence was evaluated by performing a Boltzmann average over emitting excitons for each disordered chain in the ensemble. When disorder within each chain is minimal ($\beta > 0.95$) the form of the calculated temperature dependence is similar to what was derived in Ref. [32] in the limit of no disorder. The initial increase with $T$ is a result of increasing population of the optically allowed higher energy exciton, as occurs in interchain H-aggregates. The peak occurs when $kT$ is approximately equal to the interchain splitting, and the subsequent demise is the expected behavior for J-aggregates. When disorder increases, the temperature dependence flattens out considerably and better agrees with the measured trend, as depicted in Fig. 7. Comparison of calculated and measured data show that the spatial correlation parameter β is most likely less than approximately 0.7 in both low and high molecular weight samples.

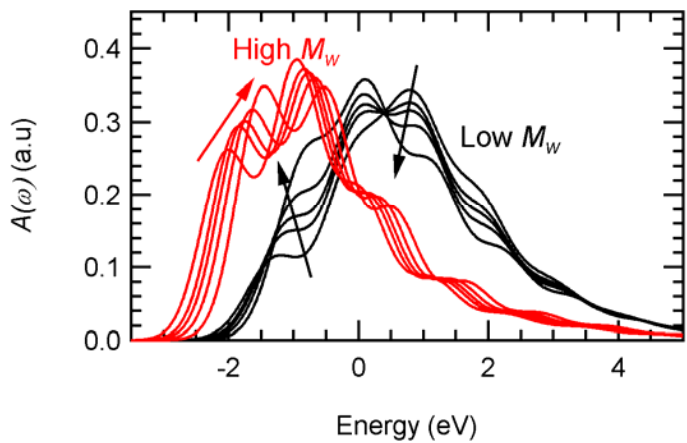

**Fig. 8** Calculated absorption spectra for the same triple points as appears in Fig. 4. Arrows indicate direction of increasing $\beta$ (0, 0.5, 0.7, 0.9, 0.999).

It is tempting to compare the measured absorption spectra for P3HT of low- and high-$M_w$ with those calculated using the triple points derived from the PL analysis. Of course this is meaningful only if one assumes that absorption and emission takes place from configurationally equivalent species, which is unlikely given the significant spectral relaxation. Fig. 8 shows calculated absorption spectra for a range of ($J_{inter}$, $J_{intra}$) points from Table 1. In this calculation we assumed homogenous broadening only, with a line width (full width at 1/e) given by $2\sigma = 1.2\omega_0$, which best describes experiment. We have checked that, unlike the PL spectra, disorder of the level studied here does not have a serious impact on the relative vibronic intensities, since the exciton coupling is too weak to induce motional narrowing. Hence, the homogeneous spectra in Fig. 8 are good approximations to ensemble-averaged spectra, and are computationally much easier to obtain. From Fig. 8 and Table 1, the values of $W = 4J_{inter}$ which give the $A_{0-0}/A_{0-1}$ ratios which best agree with experiment (see Fig. 2(a)) occur at roughly $\beta = 0.7$-0.9, and are given by W ≈ 0.08 and 0.16 eV for P3HT of high and low $M_w$, respectively. These values are somewhat larger than those predicted from the H-aggregate model. For example, in the low-$M_w$ materials, the H-aggregate model predicts $W = 0.12$ eV.

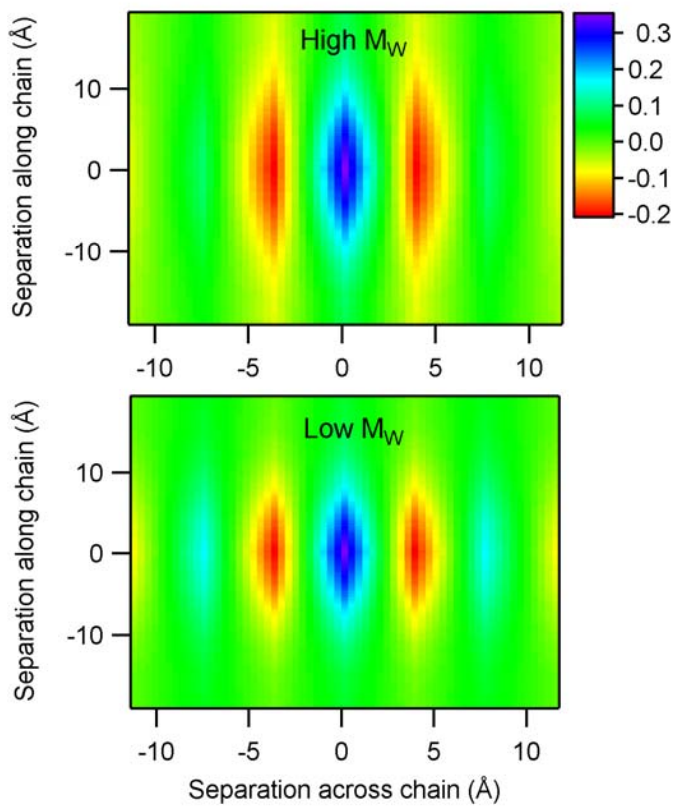

**Fig. 9** Coherence function from Eq. 13 for low- and high-$M_w$ samples and $\beta = 0.6$. Calculations are conducted on 6 by 10 aggregates (6 chains each containing 10 thiophene rings).

## C. Exciton Coherence

In what follows we will consider exciton coherence within the two-dimensional P3HT π-stacks. Generally, the coherence length is limited by the localizing influences of disorder, vibronic coupling and increasing temperature. In P3HT π-stacks we expect the coherence length along the polymer backbone, $L_{\parallel}$, and normal to the polymer backbone, $L_{\perp}$, to differ. Moreover, as we will show, both coherence lengths depend on the polymer molecular weight.

For a thermal distribution of low-energy excitons responsible for emission and transport all information regarding exciton coherence is contained in the ensemble-averaged coherence function, given by[19,43]

$$\bar{C}(\boldsymbol{r}) \equiv \left\langle \left\langle \Psi^{(em)} \Big| \sum_{\boldsymbol{R}} B_{\boldsymbol{R}}^{\dagger} B_{\boldsymbol{R}+\boldsymbol{r}} \Big| \Psi^{(em)} \right\rangle \right\rangle_{C,T}, \qquad (12)$$

where $B_{\boldsymbol{R}}^{\dagger} \equiv | \boldsymbol{R}; vac > < g; vac |$ creates an exciton at site $\boldsymbol{R}$ with no vibrational quanta ($vac$) relative to the ground state unshifted potential well. Here, the vector $\boldsymbol{R} = (n,s)$ locates the $n$th repeat unit on the $s$th chain. $\langle ... \rangle_{C,T}$ represents a dual configurational and thermal average, the former taking place over the various realizations of site-energy disorder and the latter taking place over a Boltzmann distribution of emitting excitons. The $\boldsymbol{r}$-dependent coherence in Eq. (12) is similar to that used by Mukamel and coworkers[44-46] and Kuhn and Sundstrom.[47]

Coherence in P3HT using the H-aggregate model was investigated in Ref. [19]. As the H-aggregate model is one-dimensional, only the coherence along the π-stacking direction was considered. It was shown that in *linear* H- aggregates the coherence function oscillates along the aggregate axis, changing sign as $(-1)^n$, where $n$ is the chromophore index. The oscillation reflects the dominant admixture of the high wave vector ($k=\pi$) exciton in the band-bottom excitons, as is characteristic of disordered H-aggregates.[19] By contrast, in linear J-aggregates the coherence function is nodeless, reflecting the dominant admixture of the $k = 0$ exciton in the band-bottom excitons.[37]

Fig. 9 shows the two-dimensional coherence functions for films of high- and low-$M_w$ P3HT evaluated from Eq. 12 and using $\beta$=0.6, with the coupling parameters taken from Table 1. Interestingly, the coherence function contains properties of H- and J-aggregation: the oscillations along the π-stacking direction results from the positive sign of the interchain coupling and signals H-aggregation, while the uniform phase of $\bar{C}(\boldsymbol{r})$ along the chain direction results from the negative sign of the intrachain interactions and signals J-aggregation. Fig.9 shows that the coherence has greater extent along the polymer chain for P3HT of higher $M_w$ where the intrachain coupling is about 50% larger than in the low- $M_w$ materials (see Table 1). Although not as obvious, the opposite holds for the coherence along the π-stacking axis; it has a greater extent for films prepared with low- $M_w$ materials, where the interchain coupling is twice as large compared with P3HT of higher $M_w$ (see Table 1). Fig.9 also shows that the coherence function is more anisotropic in the higher- $M_w$ films.

Based on the coherence function in Fig. 9 one can approximate the total number of molecules $N_{coh}$ within the coherence "area" defined by the spatial extent of the *envelope* of the coherence function. A more quantitative evaluation of $N_{coh}$ follows from the simple relation, [19,37]

$$N_{coh} = \bar{C}(0)^{-1} \sum_{\boldsymbol{r}} \left| \bar{C}(\boldsymbol{r}) \right|, \tag{13}$$

where the dimensionless vector $\boldsymbol{r}$ runs over all site-separation vectors within the $\pi$–stack. The absolute value dependence on $C(\boldsymbol{r})$ eliminates the phase oscillations, since $N_{coh}$ is determined by the envelope of the coherence function. For example, in a linear H-aggregate that is fully coherent (so that $N_{coh}=N$) the oscillations present in $\bar{C}(\boldsymbol{r})$ lead to complete destructive interference in the sum, $\sum_{\boldsymbol{r}} \bar{C}(\boldsymbol{r})$, making it necessary to instead sum over $|\bar{C}(\boldsymbol{r})|$ in order to determine $N_{coh}$. Indeed, the sum in Eq.(13) gives the correct result, $N_{coh}=N$ (assuming, as we are, periodic boundary conditions).

The coherence number $N_{coh}$ depends mainly on the nature of disorder ($\sigma$, $l_0$) as well as the temperature. Generally, increasing disorder and/or temperature localizes excitations and reduces $N_{coh}$ as demonstrated in detail in Refs. [19] and [37] for H and J-aggregates, respectively. Vibronic coupling is also important since it weakens the excitonic coupling, thereby allowing the disorder-induced localization to be more effective. Using Eq. (13) we find that $N_{coh}$ is 16.4 for the low-$M_w$ P3HT films of Fig. 9b and 17.9 for the high-$M_w$ films of P3HT of Fig.9a. Since the temperature is low (T = 10 K) it is disorder that is mainly responsible for localizing exciton coherence in both P3HT of low and high-$M_w$.

In the $\pi$-stack, one can also define the coherence *lengths* along the polymer chain direction ($L_\parallel$) and along the $\pi$-stacking axis ($L_\perp$) from the coherence function via,

$$L_\parallel \equiv d_\parallel \{ \bar{C}(0)^{-1} \sum_{\boldsymbol{r} \in chain} | \bar{C}(\boldsymbol{r}) | - 1 \}, \tag{14a}$$

$$L_\perp \equiv d_\perp \{ \bar{C}(0)^{-1} \sum_{\substack{\boldsymbol{r} \in chain \\ normal}} | \bar{C}(\boldsymbol{r}) | - 1 \} \tag{14b}$$

where $d_{\parallel}$ ( $d_{\perp}$ ) is the nearest neighbor distance between two adjacent thiophene units within a chain (on neighboring chains). As stated earlier we take both distances to be 4.0 Å in our calculations. Note that the term $r = 0$ is included in both sums of Eq.(14a,b). Also note that in the limit of strong delocalization where $\overline{C}(r) = \overline{C}(0)\delta_{r,0}$ the coherence lengths properly tend to zero.

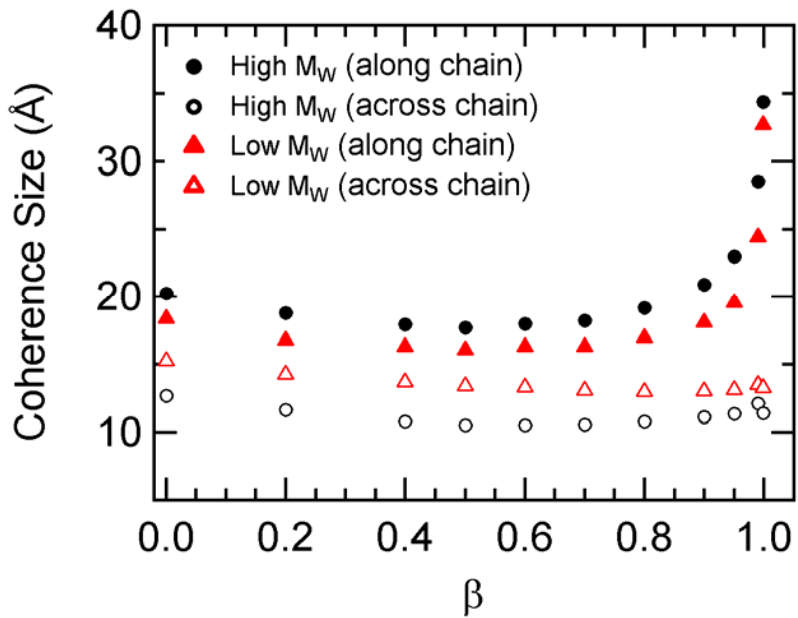

**Fig. 10** Coherence size defined in Eq.11 as a function of disorder parameter, β, for P3HT films of low and high $M_w$. Each point corresponds to triple point of table 1. Calculations are conducted on 6 by 10 aggregates (6 chains each containing 10 thiophene rings).

Based on the coherence functions in Fig. 9 and the use of Eqs. 14(a),(b), the coherence lengths along and normal to the polymer backbone can be evaluated. Fig. 10 shows the coherence lengths for both low and high- $M_w$ materials at each of the triple points of Table 1. The coherence size is remarkably stable except when $\beta$ approaches unity – at which point the correlation length becomes much larger than the aggregate dimensions. When $\beta = 1$ each repeat unit in a given aggregate has essentially the same transition frequency, i.e., the intra-aggregate disorder is zero. Since our calculations in Fig.'s 9 and 10 are based on 6 by 10 aggregates (6 chains with 10 thiophene rings per chain) the coherence length along the chain approaches $9d = 36$Å, as is observed in Fig. 10. The coherence length across the chains should approach $5d =$ 20Å, however in this limit the interchain interaction is only several times $kT$ at T = 10 K ($J_{inter}$ = 0.003 eV from Table 1) so that the thermally-induced localization remains significant.

From Fig. 10, the coherence length along the chain for P3HT of high $M_w$ is about 17 Å, which corresponds to slightly greater than four thiophene units. The coherence length drops to about 7 Å across chains. In materials of low $M_w$ the coherence is more isotropic; the coherence length along the chain is about 20% smaller than for high-$M_w$ P3HT, although the coherence length across chains is significantly higher, approximately 10 Å, which is only slightly smaller than the value of $\approx 12$ Å obtained using the H-aggregate model in Ref.[19]. The greater coherence length along the chain for P3HT of high $M_w$ is consistent with reduced torsional disorder while the larger coherence length across chains for materials of low $M_w$ arises mainly from the much higher interchain interactions ($J_{inter}$) within the π-stacks of these materials (see Table 1), possibly due to the higher torsional disorder in such structures, e.g. induced through end-group effects.

*Relationship between $N_{coh}$ and the PL line shape*

We have written extensively on the uniquely coherent nature of the 0-0 peak within the PL vibronic progression.[5,20,37] For any aggregate type (H and J) the (dimensionless) 0-0 line strength is directly related to the averaged coherence function through,

$$I_{PL}^{0-0} = \sum_{\boldsymbol{r}} \overline{C}(\boldsymbol{r}) \qquad (15)$$

In the case of J-aggregates, where $\overline{C}(\boldsymbol{r})$ is uniformly positive, the 0-0 peak is a direct measure of the coherence size and is the source of superradiance (see Eq.(13) with $|\overline{C}(\boldsymbol{r})| = \overline{C}(\boldsymbol{r})$ ). Since the side-band linestrengths are largely incoherent, the ratio $I_{PL}^{0-0} / I_{PL}^{0-1}$ is useful for probing coherence. In Ref. [37] we obtained the simple relationship,

$$I_{PL}^{0-0} / I_{PL}^{0-1} \approx N_{coh} / \lambda^2 . \qquad \text{linear J-agg} \qquad (16)$$

Eq.(15) provides a simple means of extracting $N_{coh}$ directly from the PL spectrum in J-aggregates.[31,37]

In marked contrast, in H-aggregates the 0-0 intensity (as well as ratio $I_{PL}^{0-0} / I_{PL}^{0-1}$ ) is *not*

directly proportional to the coherence number because the phase oscillations in $C(r)$ lead to destructive interferences in $I_{PL}^{0-0}$ (see Eq.(15)) but not in $N_{coh}$ (see Eq. (13)). As a result, in H-aggregates the PL ratio is *inversely* related to $N_{coh}$, decreasing with increasing $N_{coh}$ as the destructive interference between chains become more effective.[19] In the limit of maximum coherence ($T = 0$ K and $\sigma = 0$ or $l_0$=inf) $N_{coh}$ becomes equal to $N$ and the PL ratio vanishes in H-aggregates – there is no 0-0 peak by symmetry. By contrast, in the same limit the PL ratio is maximized (=$N/\lambda^2$) in J-aggregates. Hence, for H-aggregates, there is no simple relationship relating the PL ratio to $N_{coh}$. Expressions like Eq. 3 imply that $N_{coh}$ is a complex function of the exciton bandwidth, the nature of the disorder, and the vibronic coupling. However, once a model for disorder is assumed, $N_{coh}$ can be determined numerically from the measured PL ratio as was done in Ref.[19] Hence, for either aggregate type (H or J) $I_{PL}^{0-0} / I_{PL}^{0-1}$ provides a means of determining $N_{coh}$.

In the $\pi$-stacks of interest here, the PL ratio is enhanced by the coherence along the polymer chain as the transition dipoles of the repeat units are aligned in-phase. However, between chains there is a phase shift (see Fig. 10) which causes destructive interference between the chains. Hence, the H-like interchain character leads to an attenuation $I_{PL}^{0-0} / I_{PL}^{0-1}$. In the presence of disorder and thermal fluctuations the overall PL ratio is therefore due to a *competition* between intrachain, J-favoring interactions and interchain, H-favoring interactions. As we have shown here in P3HT films the competition is also a function of the chain conformation dictated by the material's molecular weight, with the most influential factor being the diminished interchain coupling (and enhanced intrachain coupling) experienced by the more planar (torsionally less disordered) macromolecules of P3HT of higher $M_w$.

## III.   DISCUSSION

From the work presented in Section III, we identify a clear microstructure-dependent interplay of intermolecular and intramolecular exciton spatial coherence in neat regioregular P3HT, which we rationalize in terms of a hybrid HJ-aggregate model. We can summarize the important conclusions as follows: (i) the photophysics of P3HT films results from a competition of intrachain $\pi$-electron coupling (J-aggregate-like behavior), and interchain Coulombic coupling (H-aggregate behavior); (ii) while the ratio of the 0-0 and 0-1 PL intensities is sensitive to interchain exciton coherence, the 0-2/0-1 ratio in the PL spectrum is predominantly an intrachain property that determines the effective Huang-Rhys factor $\lambda_{eff}^2$ for each chain in the stack (see eq. 10). The measured change in $\lambda_{eff}^2$ from near unity in P3HT samples of high $M_w$ to approximately 1.3-1.5 in low- $M_w$ materials reflects greater intrachain coupling in P3HT of high $M_w$, which we attribute to reduced torsional disorder; (iii) the absorption spectral ratios remain fairly well-described by the H-aggregate model although the HJ-aggregate model requires higher interchain interactions in order to compensate for the dominant intrachain interactions present in P3HT. The HJ-aggregate model also results in increased oscillator strength at higher energies; (iv) the enhanced 0-0/0-1 PL ratio in films of P3HT of higher $M_w$ is mainly due to decreased interchain couplings which correlate to longer conjugation lengths (and coherence lengths) within the chains. Enhanced intrachain coupling also favors a larger 0-0/0-1 PL ratio. v) for both P3HT of high and low $M_w$, the coherence length is larger along the chain than across chains, but the coherence lengths do not dramatically differ between material of low and high $M_w$.

We expect that the interplay between inter- and intrachain interactions is a general property in polymeric semiconductors. For instance, Sebastian et al. recently presented convincing evidence of a second-order phase transition in MEH-PPV dissolved in methyltetrahydrofuran,[2] where the disordered "blue-phase" is converted to the aggregate "red-phase" as the temperature is lowered through the critical temperature near 200 K. In contrast to the PL line shape of the P3HT film aggregates studied here, the PL spectrum from red-phase MEH-PPV aggregates is dominated by the 0-0 transition, indicative of J-aggregate like behavior resulting from dominant intrachain interactions. In fact, the spectra closely resemble the PL spectra obtained from the P3HT whiskers of Niles et al.,[1] the optical behavior of which were also rationalized in the same way. It is not fully understood why the optical properties of P3HT films cast from organic solvents under ambient conditions should more closely resemble H-aggregates, but the work here supports the hypothesis that such aggregates are comprised of chains with more torsional disorder, shorter conjugation lengths, and therefore stronger interchain interactions. The correlation between increased intrachain spatial coherence and reduced interchain coupling is consistent with the well known oligomer length effect: the longer the oligomer the weaker are the interchain excitonic coupling.[48-50]

The solid-state microstructures formed over the wide molecular weight range studied in this work are very distinct, ranging from a one-phase, chain extended (paraffinic-like) structure to a two-phase semicrystalline morphology,[16] and the absorbance and PL spectral lineshapes vary accordingly (Figs. 1 and 2). It is striking that while field-effect mobilities[12-16] time-of-flight mobilities,[16,17] and microwave conductivities[3] all display marked dependence on microstructure, the two-dimensional equilibrium exciton spatial coherence function derived from analysis of the PL spectral lines shape varies in a modest way over the same range (Figs. 9,10). This reflects the strong Frenkel-like character of excitons in P3HT, which, together with the strong coupling to vibrational degrees of freedom and the substantial amount of configurational disorder intrinsic in these semicrystalline microstructures, results in fairly strong localization. From Fig. 10 we obtain a coherence "area" of roughly 120 $\text{Å}^2$ for both low- and high-$M_w$ materials, with P3HT of high $M_w$ being more anisotropic ("rectangular"). We have focused mainly on the steady-state exciton coherence, although our time resolved results in Fig. 3 indicate that for time windows beyond the ~200 fs time resolution of our experiment, the spatial coherence properties do not change significantly in P3HT of high $M_w$. In contrast, the significant reduction in the PL 0-0/0-1 ratio with time over the first 5-10 ps in materials of low $M_w$ is consistent with an increasing coherence domain size with time due to planarization of the more torsionally disordered P3HT chains.[39] Interestingly, Banerji et al. reported that in films based on P3HT of $M_w$ = 79.9 kg/mol, this relaxation of the 0-0 intensity on picosecond timescales takes the system from a predominantly J-like emission spectrum (the 0-0 peak is *more* intense than the 0-1 peak), to distinctly an H-like emission spectrum (suppressed 0-0 intensity with respect to the rest of the progression),[40] which suggests that at early times the exciton has more intrachain character. In P3HT of high $M_w$, where torsional relaxation is not as significant as in low-$M_w$ materials, the situation for very early times (<< 100 fs) is quite interesting, as we expect large coherence domains before disorder-induced dephasing can occur (i.e. before disorder-induced localization) and before population (spectral) relaxation due to interactions with the bath. During such early times an exciton in a bulk heterojunction cell may be coherently connected to more electron acceptors than at steady state, greatly facilitating early-time photoinduced charge-transfer events. Such a scenario involving hot, spatially coherent excitation at organic heterojunctions in the photocarrier photogeneration mechanism in solar cells has received significant support in the last year.[40,51] Interestingly, photovoltaic efficiency performance in P3HT:PCBM diodes is found to be optimized with P3HT of medium $M_w$.[52] We suggest that the early-time spatial coherence properties of polymer films may depend on the complex configurational and energetic landscape in similar ways as the equilibrium spatial coherence properties investigated here. Finally, it is worthwhile to highlight that valuable, structural information can be obtained for a given polymer architecture based on the spectroscopic tools and analysis presented here – most prominently, the chain conformation of the constituting macromolecules. Indeed, our data implies that architectures made of materials of higher $M_w$ feature chains (or chain segments) of lower torsional disorder when compared to their low-$M_w$ counterparts, despite the fact that, e.g. their paracrystallinity[53] along the π-stack, which describes the degree of structural disorder in an imperfect crystal, is considerably higher[16,54]. Clearly, planarization of the backbone not necessarily assist π-stack formation, or put *vice versa*, high torsional disorder in macromolecular chains does not exclude good molecular packing.

## V.    CONCLUSION

In this paper we have shown how to analyze the steady-state photoluminescence spectral line shape in order to obtain the two-dimensional exciton coherence function in semicrystalline polymeric semiconductors. In particular, the 0-0/0-1 PL ratio results from a competition between intrachain interactions, which serve to enhance the PL ratio (as is characteristic of J-aggregates) and interchain interactions, which serve to attenuate the PL ratio (as is characteristic of H-aggregates). The resulting coherence function for P3HT π-stacks is J-like (nodeless) along the chain, and H-like (oscillating phase) across the chains. By contrast, the 0-2/0-1 PL ratio, quantifying the effective Huang-Rhys parameter of the vibronic progression, is sensitive primarily to intrachain excitonic coupling. In regioregular P3HT films, the interplay between intra- and intermolecular interactions results in a coherence "area" of approximately 120 square Ångstroms. Low molecular weight ($M_w$ < 50 kg/mol) materials form chain-extended morphologies based on lamellar stacks of torsionally disordered macromolecules, leading to excitons with interchain spatial coherence that extends over ~10 Å (i.e. over 2.5 neighbouring macromolecules), but with intrachain spatial coherence limited to ~12 Å (i.e. three thiophene monomers). On the opposite regime, films made of high-molecular-weight P3HT ($M_w$ > 50 kg/mol), which are characterized by two-phase morphologies consisting of interconnected crystalline and amorphous domains, host excitons that are spatially coherent over only ~7 Å , but over ~17 Å along the chain. Hence, the coherence areas for P3HT of low and high $M_w$ are similar, although high-$M_w$ materials host a more anisotropic coherence favored along the polymer chain axis. This establishes that there is a microstructure-dependent interplay between intrachain and interchain exciton interactions in this important class of semiconductor materials. Future work will be directed at exploring the time dependence of the spatial coherence as tracked by the time-resolved PL line shape.

**Acknowledgments:**

The authors dedicate this manuscript to the memory of Gianluca Latini. We acknowledge gratefully Felix Koch and Paul Smith for supply of materials and for discussions leading to the development of the polymer science reported in this paper. CS acknowledges funding from the Natural Sciences and Engineering Research Council of Canada (NSERC), the Fonds du Quebec de Recherche – Nature et Technologies (FRQNT), the Canada Research Chair in Organic Semiconductor Materials, and the Leverhulme Trust. NS is supported by a European Research Council (ERC) Starting Independent Researcher Fellowship under the grant agreement No. 279587. FS acknowledges support by the National Science Foundation under the grant No. DMR-1203811. SAH acknowledges financial support from the Engineering and Physical Science Research Council (EPSRC) through the UK-India (EP/H040218/2) program. We acknowledge support from the Royal Society through award of a University Research Fellowship (SAH). NB and MC thank the NSERC and FRQNT for the financial support, and Compute Canada for the computational ressources.